\documentclass[usenatbib]{aat} 
\begin{document}
\setcounter{page}{1}
\nametom{X(X), \pageref{firstpage}--\pageref{lastpage} (XXXX)}
\title{FUors, EXors and the role of intermediate objects}
\author{T.Yu. Magakian, T.A. Movsessian, H.R. Andreasyan}
\institute{Byurakan Astrophysical Observatory,  0213,Aragatsotn reg.,Armenia\\ \email{tigmag@sci.am}  
} 
\date{Received date}
\titlerunning{Young eruptive stars}
\authorrunning{First Author, Second Author} 

\maketitle
\label{firstpage}

\begin{abstract} 
The studies of FUors, EXors and other young eruptive stars are very important for the understanding of the earliest stages of pre-main-sequence evolution. We describe the current situation in this field. 
This is the short version of the review, presented in ``Non-stationary processes in the protoplanetary disks and their observational manifestations'' conference, Crimean Astrophysical Observatory.
\end{abstract}

\section{Introduction}

Eruptive young stellar objects (YSO) make up a negligible fraction of known YSOs. But it is their study that can be very important for understanding the early stages of the evolution of low- and medium-mass stars.
And we still do not know whether the phenomenon of eruption is a short-term, but ordinary event on this evolutionary stage, or whether it occurs only under special circumstances.

Eruptive YSO classes:
\begin{itemize}
  \item FUors
  \item FU Ori – like
  \item EXors
  \item Intermediate objects
\end{itemize}

Next, we briefly discuss the properties of these objects.

\section{FUors and FU Ori -like objects}

Four so-called \textbf{classic FUors}:
FU Ori,  V1057 Cyg, V1515 Cyg, V1735 Cyg

The outburst of FU Ori was discovered about 90 years ago; other three objects were found in the 70s. On the base of these discoveries a new distinct class of young eruptive stars was introduced \citep{Herbig1966,Ambartsumian1971}. The general features of FU Ori-like objects or simply FUors (the name suggested by Ambartsumian) were defined in the work of \citet{Herbig1977}.

\textbf{General properties of FUors in the optical and IR ranges} \citep{Herbig1977,Audard2014,CR2018}:
\begin{enumerate}
	\item Optical flare (~ 5-6 mag.) with very slow decline (or none at all)
    \item In optics, the spectrum of an F-G supergiant, without emissions, except for the P Cyg - type profiles in H$\alpha$ and some other strong lines
    \item Gradual change in spectral type with wavelength
    \item Deep CO absorption bands at 2.29 $\mu$m
    \item The bolometric luminosity after a flare is usually about 100-300L\sun
    \item They illuminate small reflection nebulae.
    \item In most cases, they are sources of collimated flows.
\end{enumerate}

\textbf{Outdated terms that can take on a new meaning}:

\textit{Pre-fuors} (i.e. progenitors of FUors). The very active YSO V1331 Cyg was one time suggested as an example of the future FUor. However, presently we know two classical T Tau type stars (CTTS), which became FUors: V1057 Cyg and V2493 Cyg.

\textit{Post-fuors}. Currently nothing is known about them. Could they be EXors?

\textit{Sub-fuors}. While there is no good definition of this term (they must be objects that are in some sense less active than FUors), we can only guess which stars they should be - perhaps EXors or MNors? 

\textbf{List of FUors}

During last 40 years the number of known FUors was growing very slowly. Certain objects were excluded from that class after further observations. The currently accepted approach introduces two sub-classes: FUors (i.e. the objects with observed outburst) and FUor-like objects (i.e. stars with the spectra and other features typical for FUors after eruption).
Current (2021-22) list, based on the atlas of \citet{CR2018} with several additions, is presented below.

FUors:
\begin{itemize}
  \item RNO 1B (V710 Cas) 
  \item V582 Aur  
  \item V883 Ori 
  \item V2775 Ori (HOPS 223)  *
  \item FU Ori
  \item V900 Mon 
  \item V960 Mon (2MASS J06593158-0405277)
  \item V1515 Cyg 
  \item V2493 Cyg (HBC 722)
  \item V2494 Cyg (HH381 IRS) 
  \item V1057 Cyg
  \item V2495 Cyg (Braid star)  *      
  \item V1735 Cyg
  \item V733 Cep (Persson's star)
  \item Gaia 18dvy 
  \item PGIR 20dci   *

\end{itemize}

FU Ori-like objects:
\begin{itemize}
  \item RNO 1C
  \item PP13 S  *
  \item L1551 IRS5 (HBC393)  *
  \item Haro 5a/6a IRS  *
  \item IRAS 05450+0019  *
  \item Z CMa 
  \item BBW 76 (V646 Pup)
  \item V371 Ser  *
  \item Pars 21
  \item IRAS 21169+6804 (CB230 A)
  \item HH 354 IRS  *
\end{itemize}

Stars, marked by asterisk (*), are visible only in IR range

Also high mass and luminosity PTF 14jg object should be mentioned, which perhaps can be considered as ``superfuor''.

\section{EXors}

The class of EXors (by the name of prototype star EX Lupi) was introduced by \citet{Herbig1989}, who defined them as T Tauri type stars with flashes of large amplitude (up to 5 mag), but short in time and repetitive. Perhaps they are the next stage of evolution after the FUors.
As was emphasized by \citet{Herbig1989}, it is impossible to distinguish them spectrally from other CTTS.

The \textbf{classical EXors}, according to \citet{Herbig1989}, \citet{Herbig2008}:

UZ Tau E,  VY Tau,  EX Lup, NY Ori, V1118 Ori (Chanal's star), V1143 Ori (Sugano's star). 
Also PV Cep and DR Tau were considered but rejected. 

\textbf{Probably important features of classical EXors}
\begin{enumerate}
  \item Flare amplitudes are comparable to FUors, but they have a much shorter duration - from several months to several years.
  \item Bolometric luminosities during an outburst  are usually on the order of 5-30 L\sun
  \item EXors are not associated with small cometary nebulae
  \item  EXors are not related to jets and outflows
\end{enumerate}

\textbf{Supposed EXors}

There is no definitive list, because the boundaries of the class itself are vague.
“The original list of EXors has changed little since 1989” - \citep{Audard2014}.
“New EXors” \citep{Lorenzetti2012} was an attempt to identify eruptive objects similar to classical EXors in the total mass of CTTS.
As a result of these and other works, the following objects were considered as EXors:
 V512 Per (SVS 13), V1180 Cas, XZ Tau (N), LDN1415 IRS, V1647 Ori, GM Cha, OO Ser, V2492 Cyg, GM Cep and again PV Cep, and V723 Car (turned out to be a massive object).
 
Most of the candidates for EXors subsequently disappeared from this list according to various criteria (in terms of luminosity, mass, character of variability, etc.); a number of them passed into the class of intermediate objects. Besides, almost all "new EXors" turned out to be associated with optical outflows and cometary nebulae.

An analysis of the evolutionary position of EXors \citep{MS2017} leads to the conclusion that almost all “new EXors” are deeply embedded in dust objects, classical EXors are somewhat older, and that only a small part of pre-main-sequence (PMS) stars are subject to EXor-type outbursts.
In recent years, a detailed study of EXors - ``EXORCISM'' program has been launched.
In the course of this program \citep{Giannini2022}, in addition to 6 classical EXors, as well as PV Cep and DR Tau stars, the following objects were subjected to detailed studies:
XZ Tau N, V350 Cep, ASASSN-13db (very low luminosity object), V1647 Ori, iPTF15afq.
Some of them undoubtedly also are the intermediate objects.
Also among the most recent finds of the probable EXor-type objects Gaia 20eae, ESO-H$\alpha$99 should be mentioned.

\section{Intermediate objects}

This class of eruptive PMS objects cannot yet be precisely defined, but it certainly exists and has a number of common properties (at least in the visible range), combining FUors and EXors characteristics in various proportions. 
\begin{enumerate}
  \item Like FUors, they are usually associated with collimated outflows and small nebulae.
 \item Outbursts last several years (between FUors and EXors)
 \item Often have emission spectra of classic TTS.
 \item Bolometric luminosities are moderate.
\end{enumerate}

The first recognized intermediate object is V1647 Ori (McNeil's Object), found in 2004 \citep{MCN2004}.

The name MNors (coined from ``McNeil's star'' and apparently premature as no definitions exist yet) was proposed by \citet{MNORS} when searching for eruptive IR variables, deeply immersed in dark clouds.
It is possible that some of 15 objects, discovered by this group, with flare duration of 1–4 years, are indeed related to FUors and EXors.
One also can add to this list infrared variable object UKIDSS-J185318.36+012454.5, recently found in BAO.

For obvious reasons it is difficult to compile a complete current list of such objects, but it will include at least 20 objects. A number of objects were identified on the base of actually observed outbursts, a few more - as the probable FUor-like stars (with a spectrum similar to the post-outburst stage, which, however, changed a few years later) – see, in particular, \citet{CR2018}.

\textbf{Intermediate objects of the first group (with unusual lightcurve)}:

\begin{itemize}
 \item V1647 Ori (McNeil’s Object)
 \item V346 Nor (HH57 IRS)
 \item OO Ser 
 \item V1180 Cas
 \item PV Cep
 \item V350 Cep
 \item V2492 Cyg
 \item V899 Mon
 \item IRAS 20390+4642 (Gaia19bey)
 \item 2MASS 22352345 + 7517076 
 \item 2MASS 08104579 - 3604310 (Gaia19ajj)
 \item HOPS 383 – an object of Class 0, visible only in IR range
 \item V1318 Cyg S – can it be a very slow FUor?

\end{itemize}

\textbf{Intermediate objects of the second group (distinguished by optical and IR spectra)}:

\begin{itemize}
 \item IRAS 06297+1021W 
 \item AR 6a (V912 Mon)
 \item AR 6b
 \item IRAS 18270-0153W
 \item IRAS 06393+0913 (can be a brown dwarf)
 \item IRAS 18341-0113S (can be a brown dwarf)

\end{itemize}

\section{New examples of unusual eruptive PMS stars}

\subsection{V1318 Cyg S (LkH$\alpha$ 225): very slow FUor?}

This emission-line star was dicovered by \citet{Herbig1960}. Before 1991 it demonstrated strong TTS type variability with occasional flares-up.

\citet{Magakian2019} found an atypical brightness increase (>5 mag) in 2015.

The object developed to HAeBe star with 750 L\sun\ in the optical range after the outburst. Its spectrum shows strong P Cyg -type absorptions, but also typical CTTS emissions.

The blue spectrum is very reminiscent of classic FUors, in red and IR ranges it is closer to Gaia 19ajj, V2492 Cyg, and V1647 Ori. Mass of the star can be as high as 8 M\sun\ \citep{Magakian2019,Hillenbrand2022}.

\subsection{PV Cep: super-EXor?}

This variable star and the connected variable nebula were discovered at BAO in 1977 \citep{GM1977}.
Shows powerful (up to 5 mag) flashes lasting for several years, the last peak in 2017. Presently is at minimum.

Has emission-rich CTTS-type spectrum. Forbidden emissions split to 4 variable components. Spectral type is estimated as G4. Bolometric luminosity reaches 100 L\sun\ at maximum, and is about 20 L\sun\ at the minimum. The accretion rate is about 10$^{-6}$ M\sun\ y$^{-1}$, which is quite high value.

In addition to high luminosity and amplitude of flares, PV Cep is a source of a powerful and extended optical and molecular outflows \citep{Andreasyan2021,Giannini2022}.

\subsection{V350 Cep: mini-FUor?}

V350 Cep is located in NGC 7129 star forming region. It was discovered in 1977 in BAO \citep{V350}.
After the rise on 4.5 magnitudes in the beginning of 70s, star remains on the same level; thus, its light curve is almost FUor-type. Two or three times the UXor-like fadings up to 2 mag were observed, then the brightness was restored.  \citep{Semkov2017}

V350 Cep  has a rich emission spectrum of the classical TTS, which does not change, spectral type is M0-M2. Can be  a source of HH flow.
Its bolometric luminosity after the outburst is quite low: 3.3 L\sun.

\subsection{V565 Mon}

This little-studied variable star illuminates triangle reflection nebula Pars 17 and produces HH outflow. It is also extremely bright mid-IR source. The bolometric luminosity of V565 Mon is high: 130 L\sun. However, it cannot be considered as a typical HAeBe star, since its spectral type is probably near G0. The only emission line (excluding the forbidden lines, probably belonging to jet) is double-peaked H$\alpha$. On the other hand, the most unusual is the presence of strong BaII absorption lines in its spectrum. Their existence suggests either the low-gravity  atmosphere (typical for FUors), either the inexplicable overabundance of barium in a very young star. All these features, summarized in the work of \citet{Andr2021}, do not allow the easy classification of this object.

\section{Conclusion}

If eruptions of YSO are indeed due to abrupt changes in the temp of accretion, their observed manifestations are very different.
We see several classes of them.

FUors: more or less well defined. Apparently very rare or short term.

EXors: the final number of discovered objects is unclear;there is also no exact definition. Perhaps indeed there are two groups of similar objects on slightly different evolutionary stage.

Intermediate objects: their great variety is no doubt. Some of them appear to push the limits of usually accepted definitions of FUors and EXors. Most interesting objects for further research.

MNors: may not yet deserve a separate class.

HAeBe stars and other massive objects: could some of them after eruptions develop into FUors?

In addition, all these manifestations can be mixed with UXor-type activity.

Young eruptive objects also can be found among certain Class I IR sources, which change brightness and become noticeable in the optical range.
\newline

This work was partly supported by RA
State Committee of Science, in the frames of the research project 21T-1C031.

\label{lastpage}
\end{document}